

\documentclass{IEEEtran}
\IEEEoverridecommandlockouts
\usepackage{latexsym}
\usepackage{multicol}
\usepackage{algorithmic,cite}
\usepackage{algorithm}
\usepackage{amssymb}
\usepackage{amsmath}
\usepackage{epsfig}
\usepackage{psfig}
\usepackage[dvips]{color}
\usepackage{epsfig}
\usepackage{psfig}
\usepackage{url}
\usepackage{array}
\usepackage{comment}
\usepackage{multirow}
\usepackage{authblk}

\def\BibTeX{{\rm B\kern-.05em{\sc i\kern-.025em b}\kern-.08em
    T\kern-.1667em\lower.7ex\hbox{E}\kern-.125emX}}
\begin{document}

{\title{TCP-aware Cross Layer Scheduling with Adaptive Modulation in IEEE 802.16 (WiMAX) Networks}}

\author[1]{Hemant Kumar Rath}
\author[2]{Abhay Karandikar}
\author[3]{Vishal Sharma}

\affil[1]{TCS Networks Lab, Bangalore 560066, India, Email:{hemant.rath@tcs.com}} 

\affil[2]{Department of Electrical Engineering, 
IIT Bombay, Mumbai 400 076, India, Email: {karandi@ee.iitb.ac.in}}

\affil[3]{Metanoia Inc., 888 Villa Street, Suite 500, Mountain View, CA, 94041-1259, USA,
Email: {v.sharma@ieee.org}}

%
%

\maketitle


\let\oldthefootnote\thefootnote
\renewcommand{\thefootnote}{\fnsymbol{footnote}}
\footnotetext[1]{This is an expanded version of the papers presented at IEEE international conference on COMmunication System softWAre 
and MiddlewaRE (COMSWARE) 2008 \cite{comsware_paper} and International Conference on Communications (ICC) 2008 \cite{icc_paper}.}
\let\thefootnote\oldthefootnote

\begin{abstract}
In this paper, we propose Transmission Control Protocol (TCP)-aware cross layer 
scheduling algorithms in a multipoint-to-point network such as the uplink of an 
IEEE 802.16 (WiMAX) network. Inadequate bandwidth allocation to a TCP flow
may lead to timeout and since TCP source drops its congestion window ($cwnd$) 
immediately after a timeout, it may affect the average throughput adversely. 
On the other hand, since the TCP source increases its $cwnd$ only linearly upon 
the availability of bandwidth, any excess assignment of bandwidth may remain 
underutilized. The proposed scheduling algorithms address this by allocating 
the resources based on $cwnd$ and TCP timeout. Moreover, since we focus on 
uplink scheduling, we consider that only {\em flow} level resource
requirement is communicated to the Base Station ($BS$) instead of
per {\em packet} information. The schedulers also take into account the wireless channel
characteristics and are thus cross layer in nature. Through exhaustive simulations, we 
demonstrate that the proposed schedulers exhibit enhanced throughput and fairness properties 
when compared to that of Round Robin (RR) scheduler under different shadowing. We 
demonstrate a gain between 3.5 \% to 15 \% in throughput and 15 \% to 25 \% in 
channel utilization over RR scheduler under different shadowing.

\end{abstract}


\section{Introduction}
Most of the current Internet applications can be broadly divided into two types: 
real-time applications and non real-time applications. Real-time applications 
require Quality of Service (QoS) guarantees in terms of minimum bandwidth 
and maximum latency from the network. Typically, these applications employ 
Universal Datagram Protocol (UDP) as the transport layer protocol. Non real-time 
Internet applications which constitute a significant percentage 
(80-90 \% applications are TCP based) employ Transmission Control Protocol (TCP) as the transport 
layer protocol. We term these applications as TCP-based applications. Unlike 
real-time applications, the TCP-based applications do not operate within the 
strict QoS guarantee framework. Instead, a TCP source adapts its rate of 
transmission based on the feedback received from the sink. In this paper, we focus
on scheduling algorithms that take into account the characteristics of a TCP flow and 
adjusts its bandwidth allocation accordingly. Specifically, we consider the
setting of IEEE 802.16 based WiMAX network \cite{802.16}. 

Due to the recent technological developments, Broadband Wireless Access (BWA)
\cite{802.16, IntelWhite} based services turn out to be advantageous than the
traditional wired services in terms of fast deployment, flexible architecture,
scalability, nomadic access and low cost. IEEE 802.16-2004 \cite{802.16}, is a
fixed BWA standard for both multipoint-to-point and mesh mode of operation. The
standard prescribes WirelessMAN-SC air interface in 10-66 GHz bands based on a
single-carrier modulation scheme and WirelessMAN-OFDM, WirelessMAN-OFDMA air
interfaces in the band of 2-11 GHz. Along with the fixed BWA, mobile BWA is also
supported through the IEEE 802.16e-2005 \cite{802.16e} amendment.

IEEE 802.16 standard does not prescribe any particular scheduling algorithm, and thus
network elements are permitted to implement their own algorithms at the
Base Station ($BS$) for both uplink and downlink. We note that the requirements of uplink and
downlink flows are different. In the downlink of IEEE 802.16, the $BS$ has
knowledge of the queues assigned to each Subscriber Station ($SS$), the arrival time of each packet and the individual channel condition of each $SS$. Hence, the $BS$ can employ a
scheduler similar to that of traditional scheduling schemes like Weighted Fair Queuing
(WFQ) \cite{WFQ}, Self-Clocked Fair Queuing (SCFQ) \cite{SCFQ}, Worst-case Fair
Weighted Fair Queuing (WF$^2$Q) \cite{W2FQ}. However, in the uplink transmission,
the $BS$ does not have packet arrival time and queue state information of $SS$s. 
Since communicating this information at {\em packet} level has significant overhead,
these scheduling algorithms are not scalable and hence not suitable for uplink.
From scalability perspective, Round Robin (RR) or its variants are suitable candidates
for uplink scheduling. 

In this paper, we consider a variant of RR scheduling. Since the channel state of
$SS$s varies randomly across the frames, a RR scheduler would result in unfairness. 
Moreover, a RR scheduler does not allocate the resources based on
TCP characteristics. If the scheduler does not assign the adequate bandwidth to a
TCP flow, then it may lead to TCP timeout. Since a TCP flow drops its congestion 
window ($cwnd$) immediately after a timeout, inadequate bandwidth
allocation and the consequent scheduling delay may affect the average
throughput adversely. On the other hand, since the TCP source increases its $cwnd$
only linearly upon the availability of bandwidth, any excess assignment of 
bandwidth may remain underutilized. The proposed scheduling algorithm addresses this
by allocating the resources based on $cwnd$ and TCP timeout. The scheduler also 
takes into account the wireless channel characteristics and is thus cross layer in nature.
Since we focus on uplink scheduling, we consider that only {\em flow} level resource
requirement is communicated to the $BS$ instead of per {\em packet} information. 
We consider a polling based approach where the $SS$s are required to communicate
their resource requirements with the $BS$ once every few frames (called polling
epoch). Since $cwnd$ size of a $SS$ does not change for one Round Trip Time ($RTT$), we
consider the polling epoch to be the minimum of $RTT$ of all flows. 

\subsection{Related Work}
Cross layer scheduling algorithms have been extensively studied in the literature
\cite{RAPPAPORT-2, berry_dsp}. Optimal algorithms can be formulated as constrained 
optimization problems within the framework of Markov Decision Processes where the objective
is to maximize a given utility subject to some QoS constraints. However, 
optimal algorithms are often computationally inefficient and several suboptimal
algorithms have been proposed \cite{nitins_book}. In this section, we review many 
such algorithms specifically proposed within the setting of IEEE 802.16  network.
Most of these algorithms have been proposed for real-time traffic with QoS
guarantees- the only exception being \cite{windowaware_letter} where the authors 
have proposed a contention based TCP-aware uplink scheduling for IEEE 802.16
network. However, in \cite{windowaware_letter}, $SS$s do not transmit any 
bandwidth (BW) requests for scheduling, instead the $BS$ measures the send rate 
of each individual flow dynamically and assigns resources based on the measured 
send rate. This kind of dynamic send rate measurement of all TCP flows at the $BS$ 
in every frame can lead to scaling problem as the $BS$ has to keep track of the 
states of all TCP flows along with the $SS$s' requirements. Moreover, the scheme 
in \cite{windowaware_letter} does not consider the time varying nature of wireless 
channel, the effect of $RTT$ variation on the requirement, and the effect of TCP 
timeouts. By assigning resources based on the send rate only, some flows might 
get starved resulting in frequent TCP congestion window ($cwnd$) drops and throughput 
degradation. On the other hand, in this paper, we propose a scheduling algorithm 
that not only takes $cwnd$ and TCP timeout into account but also the time varying wireless channel.
 
In \cite{clau_performance, clau_qos}, the authors have analyzed the QoS support 
by providing differentiated services to applications such as Voice
over IP (VoIP) and web services. They have employed Weighted Round Robin (WRR) for uplink and
Deficit Round Robin (DRR) \cite{drr_org} for downlink scheduling. In \cite{hvglobecom_06,Hem06} have proposed fair uplink scheduling schemes for Multiclass Traffic in Wi-Max. The authors have also considered delay guarantee along with fairness. Scheduling
based on dynamic weights of the IEEE 802.16 (WiMAX) flows have also been proposed in the
literature \cite{aura_qos, yang_delay_wt, minimum_wt}. In \cite{aura_qos}, the
authors determine the weights of various flows based on the ratio of average
data rate of the individual flows to the average aggregate data rate.
\cite{yang_delay_wt} determines the weights based on the size of bandwidth
requests, whereas \cite{minimum_wt} determines the weights of the individual
flows based on the minimum reserved rate. Scheduling based on the delay
requirements of Real Time Polling Service ({\it rtPS}) and Non Real Time 
Polling Service ({\it nrtPS}) have also been proposed in
the literature \cite{delay_kang_1, delay_kang_2}. In \cite{delay_kang_1}, the
authors propose a Delay Threshold Priority Queuing (DTPQ) scheduling scheme,
which determines urgency of {\it rtPS} flows based on the delay of the Head of
the Line (HoL) packet and a fixed delay threshold. The authors also
consider adaptive delay threshold-based priority queuing in
\cite{delay_kang_2}. In \cite{wang_service_flow1, wang_service_flow2}, the authors propose
Deficit Fair Priority Queue (DFPQ) scheduling algorithm.
\cite{wang_service_flow1} uses a {\it deficit counter} to maintain the maximum
allowable bandwidth for each service flow. Based on the value of the {\it
deficit counter}, it determines the priority of each flow. In
\cite{wang_service_flow2}, the authors have exploited the use of {\it deficit
counter} for inter-class scheduling in IEEE 802.16 multipoint-to-point as well as
mesh network. 

In \cite{dusit_que}, the authors propose an adaptive queue aware uplink 
bandwidth allocations scheme for {\it rtPS} and {\it nrtPS}
services. The bandwidth allocation is adjusted dynamically according to the
variations in traffic load and/or the channel quality. Researchers have 
also exploited the Opportunistic scheduling \cite{opportun} in IEEE 802.16 (WiMAX) 
networks. Though the Opportunistic scheduling improves aggregate capacity of the network, 
performance of TCP-based application is degraded due to variable rate and delay, 
leading to unfairness among the flows. 

In \cite{Point_to_multi}, the authors have proposed a Token Bank Fair Queuing
(TBFQ) \cite{TBFQ} based scheduler for the downlink flows of an IEEE 802.16
network. It considers location dependent channel errors while scheduling and
employs credit behavior of a flow to determine a priority index. Though this 
scheme provides fairness, it does not guarantee any delay while scheduling. 
\cite{qos_hou} proposes an adaptive selective Automatic Repeat reQuest based 
scheduling scheme for $nrtPS$ applications and uses an analytical model 
for parameter manipulation. Though it provides a trade-off between utilization 
and throughout, it is more suitable for the downlink scheduling in WiMAX networks,

In \cite{qos_ranga}, the authors have proposed a QoS based uplink scheduling scheme in 
IEEE 802.16d/e (WiMAX) networks. It considers end-to-end QoS, both for real-time and 
non real-time applications and proposes a hybrid uplink scheduling algorithm, which is a 
combination of Priority (P) and Earliest Due Date (E) scheduling schemes. Even though it 
improves the utilization of the radio resources, normalized throughput drops substantially 
and access delay increases exponentially as the the number of system cells increase.

In \cite{qos_hanu}, the authors have proposed a two-phased mechanism in which resource 
allocation and QoS scheduling are considered separately for OFDMA-based WiMAX networks. 
It considers system throughput optimization and QoS implementation for various types of 
traffic flows and ensures QoS by a priority-based bandwidth management scheme. 
Further, it provides admission control to provide QoS at the individual session level. 
Though this scheme provides QoS guarantee, it does not ensure fairness and high system utilization. 

In \cite{wpmc_tcp_udp}, the authors have illustrated the performance of TCP and UDP based applications through rigorous experiments conducted in an IEEE 802.16 deployed network as well as in test-beds. It has been observed that TCP applications suffer significantly as compared to UDP applications if the scheduling scheme does not consider the nature of TCP (TCP parameters). This key observation has encouraged us to work on scheduling schemes which are TCP-aware.

In this paper, we propose scheduling algorithms for the multipoint-to-point network that 
adapts its resource allocation based on TCP parameters - congestion window and timeout. 
The resource requirements are communicated during polling at {\it flow} level. 
The algorithms also exploit the wireless channel characteristics while maintaining fairness. 

\section{System Model}\label{system_model_ch5}
\begin{figure}
\centering
\scalebox{.5}{\input{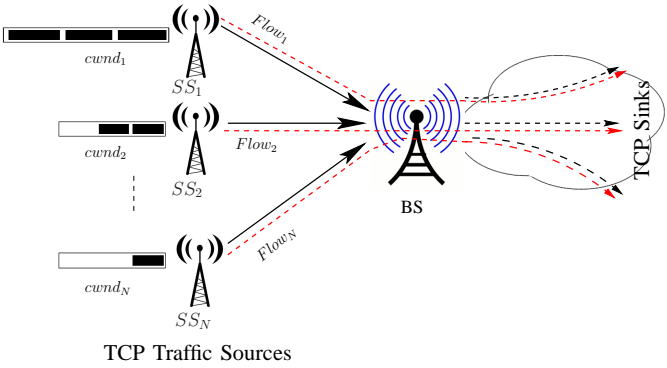}}
\caption{Multipoint-to-Point Framework with TCP-based Applications}
\label{Framework_tcp_ch1}
\end{figure}

We consider a multipoint-to-point network where multiple $SS$s are connected to
one $BS$ as shown in Fig. \ref{Framework_tcp_ch1}. This scenario may
correspond to a single cell IEEE 802.16 (WiMAX) system. 
$BS$ is the centralized entity responsible for scheduling the TCP flows. 
We assume that the $SS$s are TCP traffic sources. 
Each packet is associated with a TCP flow (source-sink pairs) and 
each flow is associated with a $SS$. Though this can be generalized to 
multiple flows per $SS$, we consider a single flow per $SS$. 
TCP acknowledgement ($ACK$) packets traverse from the sink
to the source in the downlink direction. We assume that the $ACK$
packets are very small in size and the downlink scheduler at 
the $BS$ schedules these $ACK$ packets without any delay. 

Time is divided into frames. Each frame (of duration $T_f$) in turn is
composed of a fixed number of slots of equal duration $T_s$. We assume time
varying wireless channel between a $SS$ and the $BS$. We further assume that the
channel gains between $SS$s and the $BS$ are independent and identically distributed
(i.i.d.) random variables and remain constant for a frame duration and change from frame to frame.

We assume that the individual channel state information is
available at the $BS$ in every frame. Let $SNR_i$ denote the Signal power to Noise density Ratio ($SNR$) measured
between $SS_i$ and the $BS$. Packets can be successfully received if $SNR_i$ is greater than a certain threshold
$SNR_{th}$. The value of $SNR_{th}$ depends upon the modulation and coding scheme
employed at the Physical (PHY) layer. In this paper, we consider both fixed as well as 
adaptive modulation schemes at the PHY layer.

We assume that a set $I$ of TCP flows shares a network of $I$ unidirectional
links through the $BS${\footnote{We assume one to one mapping between the flows
and the links.}}. The maximum possible data rate at link $i$ denoted by $R_i$, for
$i=1,2, 3, ... I$, is a function of $SNR$ of the corresponding link. Since the
channel state varies from frame to frame, $R_i$ also varies from frame to frame.

\section{Uplink Scheduling with Adaptive Modulation}\label{amc_tws}
Before discussing the scheduling algorithm, we define the following terms.

\begin{itemize}
\item {\it Connected Set}: The set of $SS$s that has been admitted into the
system through an admission control and connection set up phase is called
connected set ($L_{connect}$). Let $N$ be the cardinality of the connected set.

\item {\it Polling Epoch}: It is defined as the interval that the $BS$ chooses
to poll the connected $SS$s. In the proposed scheduling algorithm, the polling
is performed by the $BS$ only once after every $k$ frames.

\item {\it Schedulable Set}: A $SS$ is schedulable, if at the beginning of a
polling epoch, it has a non zero $cwnd$ and the $SNR$ of its wireless link
to the $BS$ is above a minimum threshold $SNR_{th}$. The set of such $SS$s
constitute a schedulable set $L_{sch}$. This set may change dynamically across
the polling epochs. Let $M$ be the cardinality of the schedulable set in a 
given polling epoch.

\item {\it Active Set}: A schedulable $SS$ is said to be an {\it active} $SS$ in a
frame of the polling epoch if its $SNR$ is above $SNR_{th}$ in the frame. The set of such
$SS$s constitutes an active set $L_{active}$. During a given frame of a polling epoch, the $BS$ schedules
traffic only from the active set. The membership of an active set may change dynamically across the
frames of a polling epoch, whereas the membership of a schedulable set 
changes only across the polling epochs. 

\end{itemize}

We divide the proposed scheduling algorithm into two phases: {\it polling} and
{\it slot assignment}. $BS$ polls all connected $SS$s once in every $k$ frames 
and determines the schedulable set. In each of the subsequent $k$ frames, the 
$BS$ determines the list of active $SS$s and schedules only active $SS$s on a 
frame-by-frame basis. For slot assignment, the $BS$ determines the weight of 
each active $SS$ based on the values of its $cwnd$, TCP timeout and 
accumulated deficit (as explained in the next section) and assigns slots based on its weight. At
the end of $k$ frames, the $BS$ polls the connected $SS$s again and the above process is
repeated.

\begin{figure}[h!]
\centering
\scalebox{.52}{\input{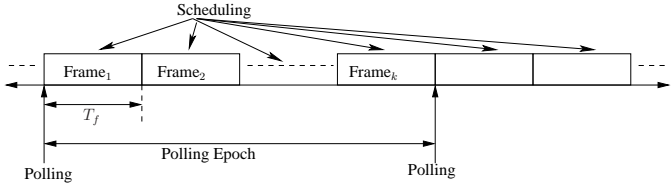}}
\caption{Polling and Frame by Frame Scheduling}
\label{sch_frame_tcp_aware_ch}
\end{figure}

The relationship between polling epoch and frame-by-frame scheduling is
illustrated in Fig. \ref{sch_frame_tcp_aware_ch}. We discuss the slot assignment
algorithm in the next section.

\subsection{Slot Allocation}\label{twus_a_sec}
Consider a polling epoch.
In the proposed algorithm, the $BS$ maintains an indicator variable $Flag_i$ 
for each $SS$; $Flag_i(n)$ is 1, if $SS_i$ is scheduled in frame $n$ of the polling
epoch. Let $N_i(n)$ be the total number of slots assigned to $SS_i$ in frame $n$. 
Let $R_i(n)$ be the rate of transmission between $SS_i$ and the $BS$ in frame $n$. 
If the underlying PHY layer employs fixed modulation scheme $R_i(n)$ is considered 
to be fixed in a polling epoch, else, it varies on a frame by frame basis.

Let $PL$ denote the length of a packet in bits. The amount of data (in bits) 
remaining to be transmitted by $SS_i$ at the beginning
of frame $n$, $D_i(n)$, is given by,

\begin{equation}
\begin{split}
D_i(0)=&cwnd_i \times PL,\\
D_i(n)=&D_i(n-1)- \\ 
         & Flag_i(n-1)\times N_i(n-1)\times R_i(n-1)\times T_s, \\
        & \forall i \in {L_{sch}}, \forall n \ge 1.
\label{demand_calc-a-tcp-ch1}
\end{split}
\end{equation}

The number of slots actually required by $SS_i$ in frame $n$ will be $\frac{1}{T_s}
\times \frac{D_i(n)}{R_i(n)}$. However, in the proposed algorithm, 
the slots are allocated in proportion to the weight of an $SS$.
Let $W_i(n)$ denote the weight of $SS_i$. After the
determination of weights, the $BS$ assigns slots to $SS_i, \forall i \in
{L_{active}}$ in frame $n$ using:

\begin{eqnarray}
\begin{split}
N_i(n)  =  &\frac{1}{T_s} \times \min \bigg\{\frac{W_i(n) \times T_f}{\sum_{j
\in L_{active}}W_j(n)},  \frac{D_i(n)}{R_i(n)}\bigg\},  \\
& \forall i \in {L_{active}}, ~~ \forall n \ge 1. 
\label{assign-1-a-tcp-ch1}
\end{split}
\end{eqnarray}

The first term in the braces of (\ref{assign-1-a-tcp-ch1}) corresponds to
the number of slots in proportion to the weight $W_i(n)$, while the second term
corresponds to the number of slots in proportion to the actual resource requirement
$D_i(n)$ of $SS_i$. By using the $\min$ function in the above equation, the $BS$
restricts the maximum number of slots assigned to any $SS$ by its requirement.
This ensures maximum slot utilization. 

We now outline the determination of weight of each $SS$ in the following subsections.

\subsubsection{Weight Determination in TCP Window-aware Uplink Scheduling with adaptive modulation (TWUS-A)}
To determine a fair allocation of slots, we define the notion of quantum size $Q$. 
The quantum size $Q(n)$ in each frame of the polling epoch corresponds to the 
number of bits transmitted per schedulable subscriber. Specifically, the quantum size
is updated as:

\begin{eqnarray}
\begin{split}
Q(0) = &\frac{R_{min}N_sT_s}{M}, \\
Q(n) = &\frac{1}{M}\sum_{i \in L_{sch}}\bigl(Flag_i(n-1) \times R_i(n-1) \\ 
       & \times N_i(n-1) \times T_s\bigr), \forall i \in L_{sch}, ~\forall n \ge 1,
\label{quantum-a-tcp-ch1}
\end{split}
\end{eqnarray}

\noindent where $N_s$ is the total number of uplink slots and $R_{min}$ is the minimum
rate of transmission among all modulation schemes. 

To keep a track of the number of slots assigned with respect to the quantum size $Q$, 
we introduce the notion of a deficit counter $DC_i$, similar to that of DRR
\cite{DRR}. The deficit counter $DC_i(n)$ is updated in every frame as:

\begin{eqnarray}
\begin{split}
DC_i(0)= & 1, \\
DC_i(n) = & DC_i(n-1) + Q(n) - \\
        &   Flag_i(n-1)\times R_i(n-1)\times N_i(n-1) \times T_s, ~ \\
        &  \forall i \in {L_{sch}}, \forall n \ge 1. 
\label{dc_calc_2-a-tcp-ch1}
\end{split}
\end{eqnarray}

From the above, we note that the deficit counter of $SS_i$ in each frame is updated by the difference of the number of bits
transmitted per schedulable subscriber and the actual number of bits transmitted by it. Thus, the deficit counter
corresponds to the accumulated credit by a schedulable $SS$. 
Since $DC_i(n)$ can take negative value, we define the scaled deficit counter $dc_i$ as follows:

\begin{eqnarray}
\begin{split}
dc_i(0)=& 1, ~\forall i \in {L_{active}}, \\
dc_i(n)=& DC_i(n)+|\min_j DC_j(n)|, \\ 
         & ~~\forall i,j \in {L_{active}}, ~\forall n \ge 1.
\label{scaled_dc_i-tcp-ch1}
\end{split}
\end{eqnarray}

The weight of a $SS$ is then determined in proportion to not only its resource
requirements in terms of the number of slots as indicated by $\frac{D_i(n)}{R_i(n)}$, but also to the accumulated credit in terms of the number of slots as indicated by $\frac{dc_i(n)}{R_i(n)}$, i.e., the weight $W_i(n)$ is determined using 

\begin{equation}
W_i(n)=\frac{\frac{D_i(n)}{R_i(n)} \times \frac{dc_i(n)}{R_i(n)}}{\sum_{j \in
L_{active}}\frac{D_j(n)}{R_j(n)} \times \frac{dc_j(n)}{R_j(n)}}, \forall i  \in
{L_{active}}, ~~ \forall n \ge 1.
\label{formu-wghts-1-a-tcp-ch1}
\end{equation}

The slots are then assigned using (\ref{assign-1-a-tcp-ch1}). Inclusion of transmission rate $R_i(n)$ in weight computation ensures in providing fair opportunity for the amount of data transmissions to each user, irrespective of its channel quality and transmission rate. In the next section, we incorporate TCP timeout information along with the
congestion window size to determine the weight.

\subsubsection{Weight Determination in Deadline based TCP Window-aware Uplink Scheduling with adaptive modulation (DTWUS-A)}\label{dtwus_a_sec}
The basic idea in determining the weight is that an active $SS$ whose TCP flow 
is approaching TCP timeout should be given a higher weight. Let, $TTO_i$ denote 
the time left to reach TCP timeout of $SS_i$ at the beginning of a polling epoch. 
Note that the maximum value of $TTO_i$ is the TCP timeout\footnote{TCP flows 
generally start at random and hence different flows have different residual 
times to reach TCP timeout.} associated with the TCP flow of $SS_i$. 

For each schedulable $SS$, we define {\it deadline} $d_i$ to indicate the 
urgency of scheduling. At the beginning of a polling epoch, $d_i$
of $SS_i$ is initialized to $TTO_i$. If $SS_i$ is scheduled in frame $n$, then 
$d_i(n)$ remains unchanged, i.e, it takes the value of $d_i(n-1)$. Otherwise,
$d_i(n)$ is decremented by one frame duration from its previous value. $BS$
updates the deadlines of the schedulable flows as follows:

\begin{eqnarray}
\begin{split}
\label{formu-delay-tcp-ch1}
d_i(0)= & TTO_i, ~ \forall i \in L_{connected}, \\
d_i(n)= & d_i(n-1)-T_f, ~ \forall i \in (L_{sch} \setminus L_{active}), ~\forall
n \ge 1, \\
d_i(n)= & d_i(n-1), ~ \forall i \in L_{active}, ~\forall n \ge 1.
\end{split}
\end{eqnarray}

If $T_f$ exceeds $d_i(n)$, then the deadline $d_i(n)$ of $SS_i$ is initialized
to TCP timeout ($TTO_i$) of that $SS$. In that case, TCP flow experiences a
timeout before getting scheduled, resulting in reduction of $cwnd_i$ to one.
Thus, the $BS$ incorporates the urgency measure $d_i(n)$ in computing weight $W_i(n)$ for $SS_i$ as: 

\begin{eqnarray}
\begin{split}
W_i(n)=&\frac{\frac{D_i(n)}{R_i(n)} \times \frac{dc_i(n)}{R_i(n)}/d_i(n)}{\sum_{j
\in L_{active}}\frac{D_j(n)}{R_j(n)}\times \frac{dc_j(n)}{R_j(n)}/d_j(n)}, \\
& ~\forall i \in {L_{active}}, ~~ \forall n \ge 1.
\label{formu-wghts-2-a-tcp-ch1}
\end{split}
\end{eqnarray}

Note that (\ref{formu-wghts-2-a-tcp-ch1}) is similar to (\ref{formu-wghts-1-a-tcp-ch1}) except
for incorporating $d_i(n)$. The use of the deadline in weight determination 
ensures that the weight of a $SS$ with a smaller deadline is higher as 
compared to that of another $SS$ which has a larger deadline. After the 
determination of weights, the number of slots
assigned to $SS_i$, $ \forall i \in {L_{active}}$ in frame $n$ is determined
using (\ref{assign-1-a-tcp-ch1}). 

The pseudo-code of the proposed schedulers TWUS-A and DTWUS-A is presented in
Algorithm 1. We have combined both schedulers by using
$Flag_{deadline}$, which is set to one for DTWUS-A and is set to zero for
TWUS-A.

\begin{algorithm}[h!]
\footnotesize
\label{SuperAlg3-tcp-ch1}
\begin{algorithmic}[1]\scriptsize
\caption{:TCP-aware Uplink Scheduler with Adaptive Modulation}
\WHILE {TRUE}
\STATE Determine $L_{sch}$ for the current polling epoch
\STATE $Flag_i(0) \leftarrow 0 \; \forall i  \in L_{sch}$  
\STATE $D_i(n) \leftarrow {cwnd_i} \times PL, ~ DC_i (0) \leftarrow 1, ~ dc_i(0)
\leftarrow 1, ~ W_i(0) \leftarrow 0, ~  N_i(0) \leftarrow 0   \; \forall i \in
L_{sch}$
\IF   {$Shceduler_{Type} = TWUS-A$}
\STATE $Flag_{deadline}=0$, ~ $d_i(0) \leftarrow 1\; \forall i  \in L_{sch}$
\ELSE
\STATE $Flag_{deadline}=1$, ~ $d_i(0)\leftarrow TTO_i\; \forall i  \in L_{sch}$
\ENDIF
\STATE $M \leftarrow |L_{sch}|$
\IF    {$n =1 $}
\STATE $Q(0) \leftarrow \frac{R_{min} \times N_s \times T_s}{M}$
\ENDIF
\STATE $k \leftarrow \min_i\{RTT_i\}, ~  T \leftarrow kT_f$
\STATE Frame number $n \leftarrow 1$
\WHILE    {$T>0$} 
\STATE $L_{active} \leftarrow \phi$
\FORALL {$i \in L_{sch}$}
\IF {($SNR_i(n) \ge SNR_{th}) ~ \Lambda (D_i(n-1) >1$)} 
\STATE $L_{active} \leftarrow L_{active} \cup \{i\}$
\STATE $DC_i(n) \leftarrow DC_i(n-1) + Q(n-1) - R_i(n-1)\times N_i(n-1) \times
T_s$
\IF {$Flag_{deadline}$ = 1}
\STATE $d_i(n) \leftarrow d_i(n-1)$
\ELSE 
\STATE $d_i(n) \leftarrow 1$
\ENDIF
\ELSE 
\STATE $R_i(n) \leftarrow 0, ~ D_i(n) \leftarrow D_i(n-1), ~ DC_i(n) \leftarrow
DC_i(n-1) + Q(n-1)$
\IF {$Flag_{deadline}$ = 1}
\STATE $d_i(n) \leftarrow d_i(n-1)- T_f$
\ELSE 
\STATE $d_i(n) \leftarrow 1$
\ENDIF
\IF {$d_i(n) \le 0$} 
\STATE $d_i(n) \leftarrow TO_i$
\ENDIF
\STATE $W_i(n) \leftarrow 0, ~ N_i(n) \leftarrow 0$
\ENDIF
\ENDFOR
\FORALL {$i \in L_{active}$}
\STATE $D_i(n) \leftarrow D_i(n-1) - N_i(n-1) \times R_i(n-1) \times T_s$
\STATE $dc_i(n) \leftarrow DC_i(n) + |\min_j DC_j(n)|, \forall j \in L_{active}$
\STATE Map $R_i(n)$ to $SNR_i(n)$ in Table \ref{table_multi_mod-tcp-ch1}
\STATE $W_i(n) \leftarrow \frac{\frac{D_i(n)}{R_i(n)} \times
\frac{dc_i(n)}{R_i(n)}/d_i(n)}{\sum_{j \in L_{active}}(\frac{D_j(n)}{R_j(n)}
\times \frac{dc_j(n)}{R_j(n)}/d_j(n))}$
\STATE $N_i(n)  \leftarrow \frac{1}{T_s} \times \min ~\ \bigg(\frac{W_i(n)
\times T_f}{\sum_{j \in L_{active}}W_j(n)}, \frac{D_i(n)}{R_i(n)}\bigg)$
\STATE $Q(n) \leftarrow \frac{1}{M}\sum_{i \in L_{sch}}{R_i(n-1) \times N_i(n-1)
\times T_s}$
\ENDFOR
\STATE $T \leftarrow T-T_f, ~ n \leftarrow n+1$
\ENDWHILE
\ENDWHILE
\end{algorithmic}
\end{algorithm}

\section{Implementation of TCP-aware Scheduling}\label{impli_tcp_aware_ch5}
We consider Time Division Duplex (TDD) based IEEE 802.16 (WiMAX) network, in which each frame of duration $T_f$
is divided into uplink and downlink subframes of durations $T_{ul}$ and $T_{dl}$
respectively. We consider adaptive modulation scheme at the PHY layer and employ
Quadrature Phase Shift Keying (QPSK) modulation, 16-Quadrature Amplitude Modulation 
(QAM) and 64-QAM schemes. Let $B$ denote the channel bandwidth. 
The maximum data rate ($R$) attainable for 
an Additive White Gaussian Noise (AWGN) channel can be expressed as: 

\begin{equation}
\label{new_eqn_1}
R=B \times \log_2(1+MI \times SNR),
\end{equation}

\noindent where $MI$ is the modulation index, which depends upon the desired 
Bit Error Rate (BER) and spectral efficiency of the modulation scheme. 
As discussed in \cite{ANDREA_GOLD}, for a target BER $p_b$ and spectral 
efficiency $\frac{R}{B}$, modulation index can be expressed as:

\begin{equation}
\label{new_eqn_2}
MI =\begin{cases} 
       - \frac{\ln(5 \times p_b)}{1.5}, &~\text{if} ~ \frac{R}{B} < 4, \\
      - \frac{\ln(0.5 \times p_b)}{1.5},& ~\text{if} ~ \frac{R}{B} \ge 4.
	\end{cases}
\end{equation}

Using (\ref{new_eqn_1}) and (\ref{new_eqn_2}), we determine the minimum $SNR$ 
required ($SBR_{th}$) as:

\begin{equation}
\begin{split}
SNR_{th} &= \frac{2^{\frac{R}{B}}-1}{MI} \\
& =  \frac{(1-2^{\frac{R}{B}})\times \ln(5 \times p_b)}{1.5},~\text{if} ~ \frac{R}{B} <
4 \\
& =  \frac{(1-2^{\frac{R}{B}})\times \ln(0.5 \times p_b)}{1.5},~\text{if} ~ \frac{R}{B}
\ge 4,
\label{max_datarate_2-tcp-ch1} 
\end{split}
\end{equation}


For target BERs of $10^{-5}$ and $10^{-6}$, a channel bandwidth ($B$) of 25 MHz, 
and for the data rates of 40, 80 and 120 Mbps (for QPSK, 16-QAM and 64-QAM modulation 
schemes respectively), we determine $SNR_{th}$ using (\ref{max_datarate_2-tcp-ch1}). 
These values are given in Table \ref{table_multi_mod-tcp-ch1}.

\begin{table}[h!]
\renewcommand{\arraystretch}{1}\addtolength{\tabcolsep}{-1pt}
\caption{Modulation Schemes in the Uplink of WirelessMAN-SC IEEE 802.16 (Channel
Bandwidth ~ $B= 25$ MHz)} 
\centering
\newcommand{\mc}[2]{\multicolumn{#1}{#2}}
\begin{tabular}{|l|l|l|l|l|}\hline
\mc{1}{|c|}{\bf{Modulation} } & \mc{1}{|c|}{\bf{Data Rate}} &
\mc{1}{|c|}{\bf{$\frac{R}{B}$}} & \mc{1}{|c|}{\bf{$SNR_{th}$ (dB)}} &
\mc{1}{|c|}{\bf{$SNR_{th}$ (dB)}}  \\ 
\mc{1}{|c|}{\bf{Scheme}} & \mc{1}{|c|}{\bf{$R$ (Mbps)}} &
\mc{1}{|c|}{\bf{(bps/Hz)}} & \mc{1}{|c|}{\bf{$BER=10^{-5}$}} &
\mc{1}{|c|}{$BER=10^{-6}$}  \\ \hline
\mc{1}{|c|}{QPSK} & \mc{1}{|c|}{40} & \mc{1}{|c|}{1.6} & \mc{1}{|c|}{11.27} &
\mc{1}{|c|}{12.18} \\ \hline
\mc{1}{|c|}{16-QAM} & \mc{1}{|c|}{80} & \mc{1}{|c|}{3.2} & \mc{1}{|c|}{17.33} &
\mc{1}{|c|}{18.23} \\ \hline
\mc{1}{|c|}{64-QAM} & \mc{1}{|c|}{120}& \mc{1}{|c|}{4.8} & \mc{1}{|c|}{23.39} &
\mc{1}{|c|}{24.14} \\ \hline
\end{tabular}
\label{table_multi_mod-tcp-ch1}
\end{table}

In the proposed scheme, $SS$s are required to maintain a queue (per flow) at
their interfaces. Packets residing in the queue of a $SS$ is served in a
first-come first-serve basis. We assume that the $BS$ has the channel state 
information of each connected $SS$. This information is used by the
$BS$ to determine the schedulable set at the beginning of a polling epoch and to update
the active set in every frame. At the beginning of every polling epoch, each
$SS$ conveys its requirements in terms of its current congestion window
($cwnd_i$) size and time left to reach TCP timeout ($TTO_i$) to the $BS$. $BS$
in turn, determines the number of slots to be assigned based on the resource
requirement, deficit counter and deadline values of each schedulable $SS$ and
determines the modulation scheme to be used on a frame by frame basis. $BS$ conveys
this information to each $SS$ through the uplink map ($UL_{MAP}$) \cite{802.16}.

$BS$ also determines the polling epoch $k$. 
Since $cwnd$ of each flow remains fixed for one $RTT$\footnote{Typical 
TCP $RTT$s are in the range of 100 msec - 200 msec,
whereas the frame length $T_f$ in IEEE 802.16 is 0.5 msec, 1 msec or 2 msec.}, the overall resource 
requirement of each $SS$ also remains fixed for one $RTT$. Hence, the $BS$ should 
choose a polling epoch of the order of one $RTT$. If it polls 
once per multiple $RTT$s (more than one), then there is a chance of TCP timeout resulting in 
$cwnd$ reduction. In this paper, we choose polling epoch $k$ to be the minimum
$RTT$ among all the flows. This enables each $SS$ to convey 
its resource requirement at-least once in every $RTT$.

\begin{figure}[h]
\centering
\scalebox{.46}{\input{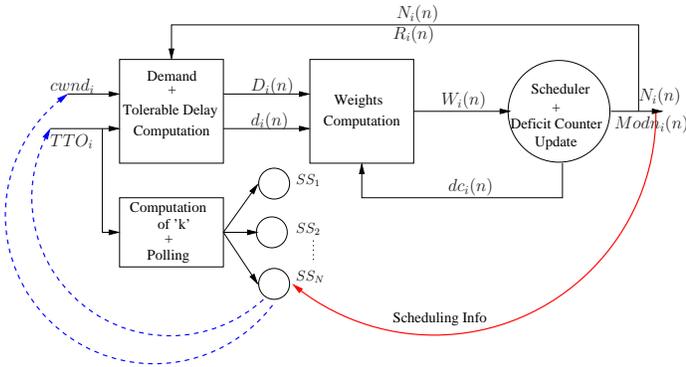}}
\caption{Block Diagram of TCP-aware Uplink Scheduler} \label{amc_block-tcp-ch1}
\end{figure}

The block diagram of the proposed uplink scheduler is shown in Fig. \ref{amc_block-tcp-ch1}.

\section{Experimental Evaluation of TCP-aware Schedulers with Adaptive Modulation}\label{sim_tcp_f_a}
In this section, we describe simulation experiments that have been performed to
evaluate TCP-aware scheduling. All the simulations have been conducted using
implementation of TCP-aware scheduling within IEEE 802.16 (WiMAX) setting in MATLAB
\cite{MATLAB}. We consider a multipoint-to-point WiMAX network where
10 $SS$s are connected to a centralized $BS$ as shown in Fig.
\ref{Framework_tcp_ch1}. We simulate one TCP flow per $SS$. Each TCP flow starts
randomly. The $RTT$s of the flows are updated using exponential averaging. Each
$SS$ is assumed to have a large enough buffer at its interface, such that the
probability of buffer overflow is negligible. The frame duration $T_f$ is set
equal to 2 msec\footnote{Frame duration ($T_f$) is equally divided between
uplink subframe ($T_{ul}$) and downlink subframe ($T_{dl}$).}. The uplink
subframe $T_{ul}$ consists of 500 data slots (we assume that the number of
control slots used is negligible). We consider both equal and unequal distances
between $SS$s and the $BS$. For equal distances, the distances of all $SS$s from
the $BS$ are 1 km each and for unequal distances, the distances between $SS$s
($SS_1$ - $SS_{10}$) and the $BS$ are 0.857 km, 1.071 km, 0.910 km, 1.230 km,
1.113 km, 0.956 km, 1.122 km, 0.884 km, 0.970 km and 1.216 km respectively.

We consider $BER=10^{-6}$ for the applications
and use the $SNR_{th}$s for selecting an appropriate modulation scheme as shown
in Table \ref{table_multi_mod-tcp-ch1}. The path loss exponent due to distance is set as $\gamma=4$.
We simulate both shadowing as well as fast fading in our experiments. We also
consider AWGN with Power Spectral Density (PSD) $N_0$ = 0.35 (4.5 dB/Hz). The shadowing is modeled as
Log-normal with mean zero and standard deviation ($\sigma$) of 8 dB. In each
simulation run, the channel gain due to Log-normal shadowing is kept fixed for a
duration of 50 frames. For fast fading, we consider Rayleigh fading model. The
channel gain due to fast fading is modeled as complex Gaussian random variable
or equivalently the power gain is an exponential random variable with mean
$\beta$. The coherence time of the channel is considered to be equal to one
frame duration, i.e, the channel gain due to fast fading changes from frame to
frame. The values of $\beta$ and transmission power are chosen such that the
expected $SNR$ received at the cell edge is more than $SNR_{th}$ required for
transmission. We also repeat the experiments with different 
values of $\sigma$--4, 6, 8, 10 and 12 dB.

We conduct eight sets of experiments based on distance (equal and unequal) and
the proposed schedulers with fixed and adaptive modulations. Note that 
TWUS and DTWUS are the fixed modulation versions of TWUS-A and DTWUS-A. In 
TWUS and DTWUS, we employ QPSK modulation scheme only, whereas in TWUS-A and DTWUS-A, 
we adapt the modulation scheme as shown in Table \ref{table_multi_mod-tcp-ch1}. 
The system parameters used for simulations are presented in Table 
\ref{system_parameters-expt-a}. The value of each performance parameter 
observed has been averaged over 50 independent simulation runs, with the ”warm up” frames
(approximately 200 frames) being discarded in each run, to ensure that the values observed were steady-state values. 
We also implement Round Robin (RR) scheduler with both fixed and adaptive modulation 
schemes and compare the performance of TCP-aware schedulers with that of RR 
schedulers. 

\begin{table}[h!]
\centering
\caption{Summary of System Parameters}
\begin{tabular}{|c|c|} \hline
\bf{Simulation Parameter} 	& \bf{Value} \\ \hline
Channel Bandwidth 		& 25 MHz \\ \hline
Adaptive Modulation Schemes  	& QPSK, 16-QAM, 64-QAM \\ \hline
Bit Error Rate 			& $10^{-6}$ \\ \hline
Path Loss Exponent ($\gamma$) 	& 4 \\ \hline
Frame Length $T_f$ 		& 2 msec \\ \hline
Uplink/Downlink Frame Length 	& 1 msec \\ \hline
Number of Data Slots per $T_{ul}$ & 500 \\ \hline
Number of Frames Simulated 	& 40000 \\ \hline
TCP Type 			& TCP Reno \\ \hline
Number of Independent Runs 	& 50 \\ \hline
Number of $SS$s 		& 10 \\ \hline
Packet Size 			& 8000 bits \\ \hline
\end{tabular}
\label{system_parameters-expt-a}
\end{table}

\subsection{Simulation Results}
\subsubsection{Impact of $cwnd_{Max}$}
\begin{figure}[h]
\centering
\includegraphics[width=0.53\textwidth]{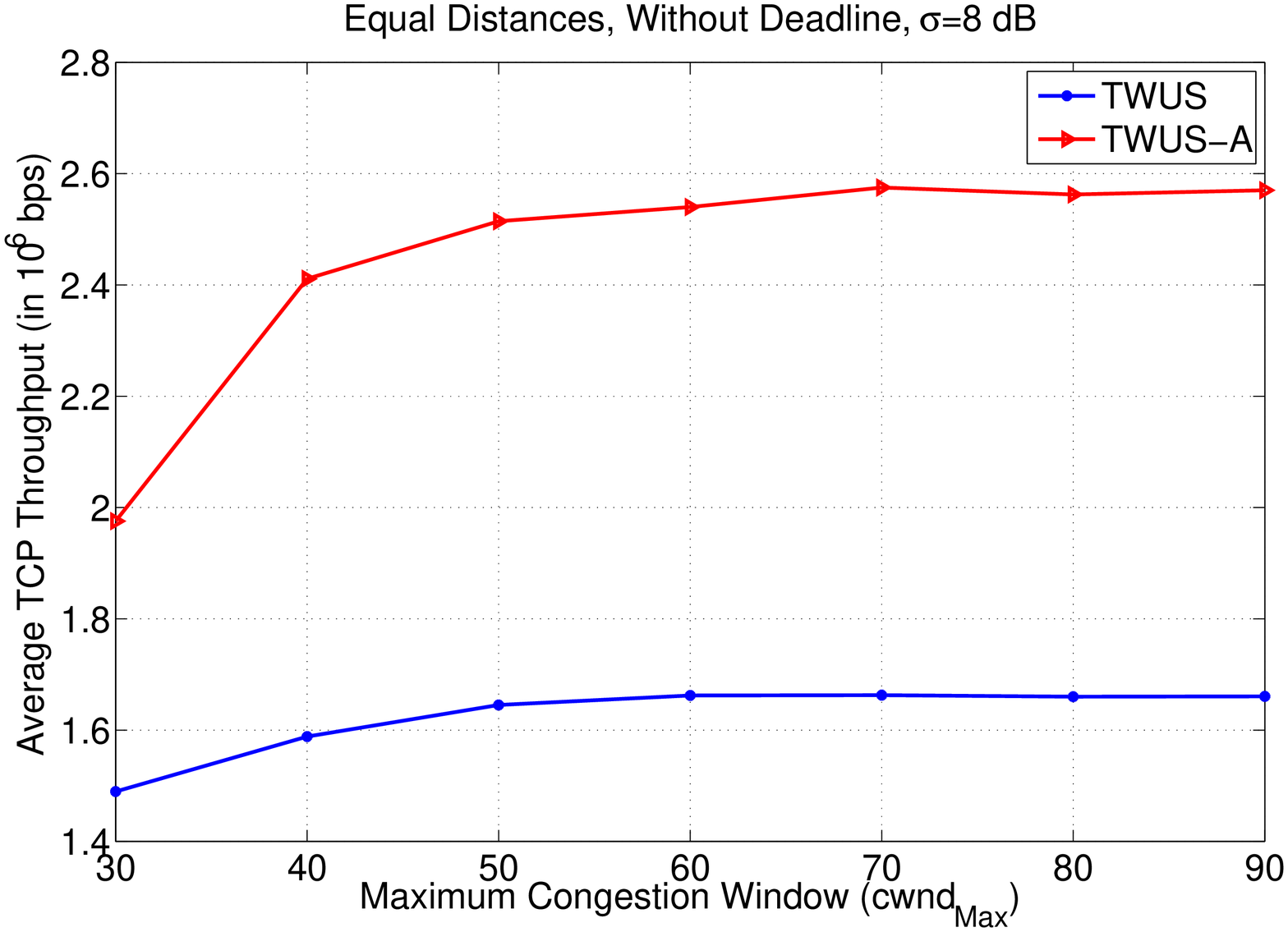}
\caption{Average TCP Throughput vs. $cwnd_{Max}$} \label{cwnd_max-expt}
\end{figure}

Since $cwnd_{Max}$ value controls the TCP throughput, choosing its correct 
value in simulations is very important. A very high $cwnd_{Max}$ will 
cause more congestion and packet drops due to buffer overflow, whereas 
a small $cwnd_{Max}$ will under-utilize the network. The value of $cwnd_{Max}$ 
should be selected depending upon the PHY layer capacity, such that 
buffer overflow is minimized and network is appropriately utilized. Therefore, 
before conducting the experiments to verify the performance of the proposed 
schedulers, we perform experiments to determine the value of $cwnd_{Max}$ at which the TCP
throughput saturates and plot the results Fig..\ref{cwnd_max-expt}. For
completeness, we plot the results of both TWUS and TWUS-A with equal distances
in this figure. From this figure, we observe that TCP throughput remains
constant (reaches saturation) once the $cwnd_{Max}$ reaches 70 packets for
TWUS-A, and 60 packets for TWUS. We choose $cwnd_{Max}$ = 70 and 60 packets for the
TCP-aware schedulers with adaptive modulation and fixed modulation schemes 
respectively in the rest of our experiments.

\subsection{Comparison with Round Robin Scheduler}
To compare the performance of TCP-aware schedulers with that of RR schedulers, 
we determine the average $cwnd$ size, average TCP throughput, slot utilization and 
Jain's Fairness Index (JFI) \cite{JAIN} achieved by each of the schedulers.

\subsubsection{Average $cwnd$ size and Throughput Comparison}
\begin{figure}[h!]
\centering
\includegraphics[width=0.53\textwidth]{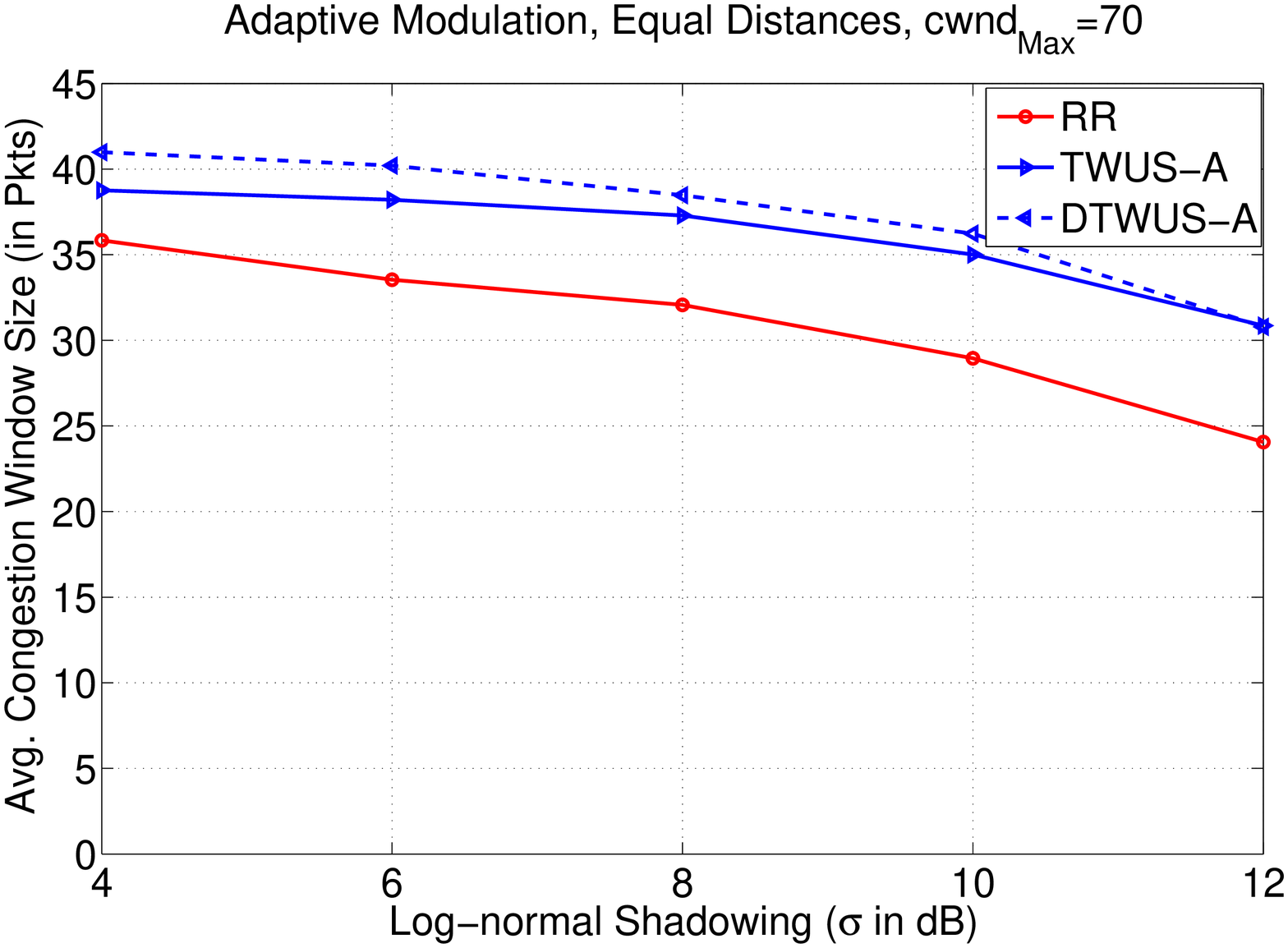}
\caption{Avg. $cwnd$ for TCP-Aware Schedulers vs. RR Scheduler}  
\label{rr-tcp-aware-comp-2}
\end{figure}

\begin{figure}[h!]
\centering
\includegraphics[width=0.53\textwidth]{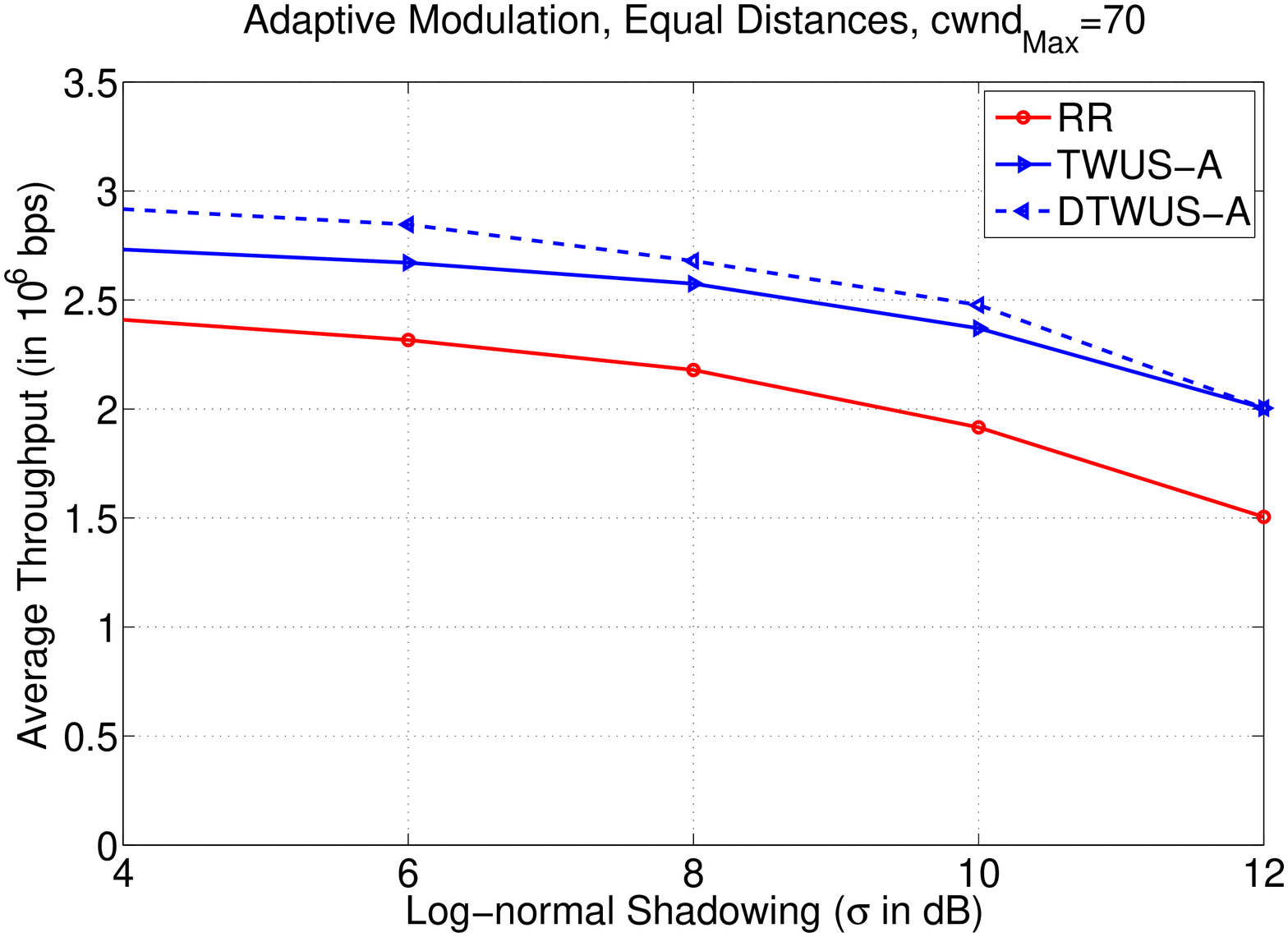}
\caption{Avg. Throughput for TCP-Aware Schedulers vs. RR Scheduler}  
\label{rr-tcp-aware-comp-4}
\end{figure}

\begin{figure}[h!]
\centering
\includegraphics[width=0.53\textwidth]{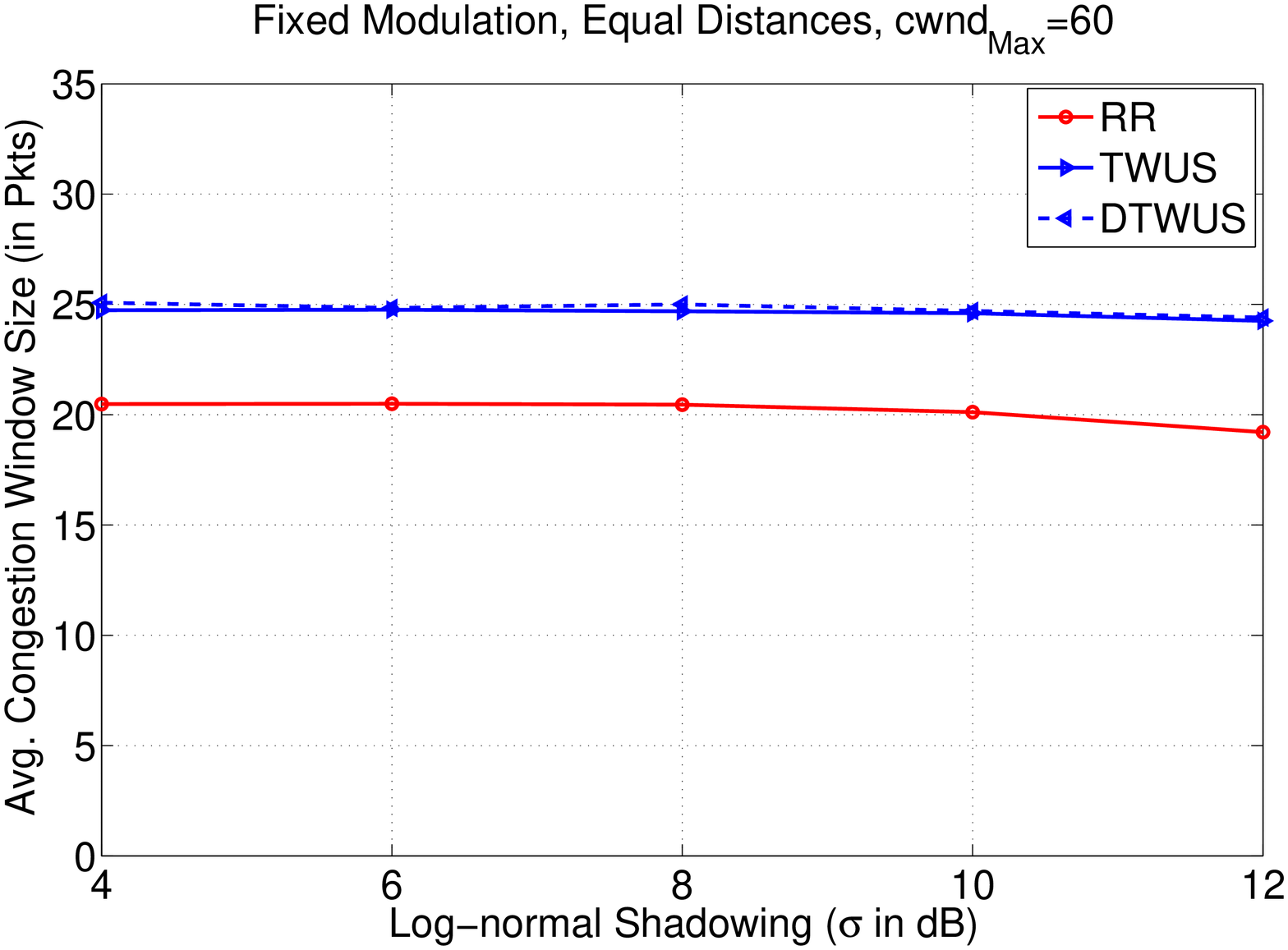}
\caption{Avg. $cwnd$ for TCP-Aware Schedulers vs. RR Scheduler}  
\label{rr-tcp-aware-comp-1}
\end{figure}

\begin{figure}[h!]
\centering
\includegraphics[width=0.53\textwidth]{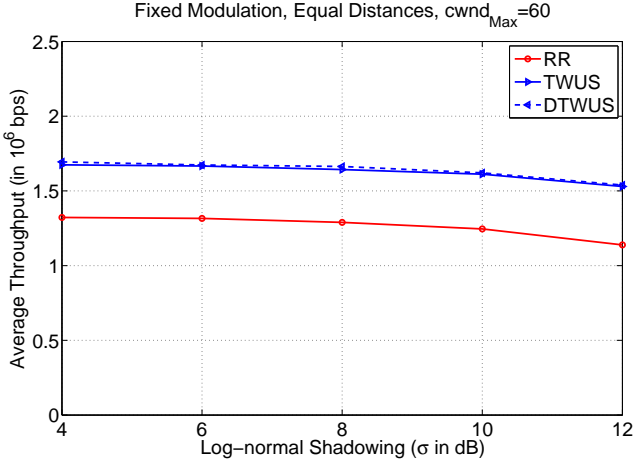}
\caption{Avg. Throughput for TCP-Aware Schedulers vs. RR Scheduler}  
\label{rr-tcp-aware-comp-3}
\end{figure}

In Fig. \ref{rr-tcp-aware-comp-2} and \ref{rr-tcp-aware-comp-4} 
(Fig. \ref{rr-tcp-aware-comp-1} and \ref{rr-tcp-aware-comp-3}), we plot the average
$cwnd$ size and TCP throughput respectively under different shadowing
with adaptive (fixed) modulation in consideration. From these figures, we observe that
the average $cwnd$ size as well as TCP throughput achieved by the TCP-aware
schedulers are higher than that of RR schedulers under different standard 
deviation ($\sigma$) of Log-normal shadowing. Moreover, the average $cwnd$ and throughput are higher in case
of adaptive modulation. We also observe that 
as the $\sigma$ of Log-normal shadowing increases, the average $cwnd$ 
size as well as TCP throughput achieved by both RR and TCP-aware schedulers decrease. 
However, the rate of decrease of $cwnd$ and TCP throughput is more in adaptive modulation 
than that in fixed modulation. Moreover, the gain in $cwnd$ size of TWUS-A over RR 
varies between 5.5\% to 16.5\%, whereas the gain in TCP throughput varies between 3.5\% to 16.5\% 
for equal distance experiments. 

Though we have illustrated the results for equal distance
experiments, similar comparisons are also valid for unequal distance experiments and for 
fixed modulation experiments.

\subsubsection{Jain's Fairness Comparison}
\begin{figure}[h!]
\centering
\includegraphics[width=0.53\textwidth]{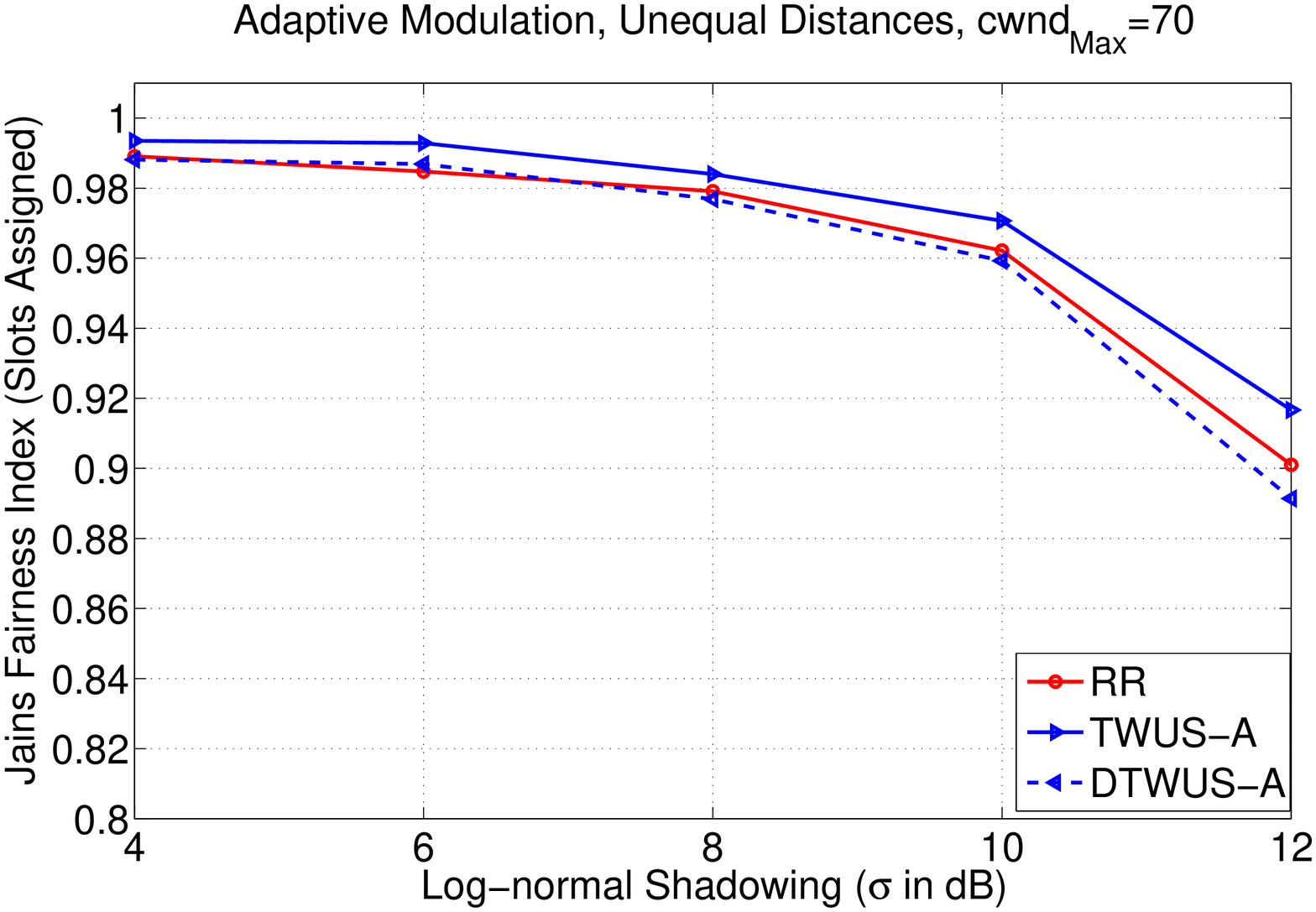}
\caption{Jain's Fairness for TCP-Aware Schedulers vs. RR Scheduler}  \label{pf-tcp-aware-jfi-comp1}
\end{figure}

In Fig. \ref{pf-tcp-aware-jfi-comp1}, we plot the variation of Jain's Fairness Index (JFI) for 
the number of slots assigned to each $SS$ for the TCP-aware and RR schedulers 
with different values of $\sigma$ of Log-normal shadowing. From this figure, we
observe that both TWUS-A and DTWUS-A have JFI above 90\% for most of the channel conditions. 
We also observe that TWUS-A is more fair than that of DTWUS-A. It is due to the fact that inclusion of deadline in 
the scheduling process reduces the fairness index. Since RR scheduler does not depend upon the demand and deadline, fairness of RR scheduler is better as compared to DTWUS-A. Similar results are also valid for equal distance experiments and 
for fixed modulation experiments.

\subsubsection{Transport Layer Fairness}
We also measure two Transport layer fairness indices, namely Worst Case TCP
Fairness Index (WCTFI) and TCP Fairness Index (TFI) as defined in \cite{TCP-fair}. TFI captures the relative fairness among the users. 
Both WCTFI and TFI are measured for the proposed schedulers with respect to Round Robin (RR) scheduler as follows:

\vspace{-1mm}
\begin{equation}
\label{wctfi-fm}
\text{WCTFI}=\min_{\forall i}\bigg[\mathcal{M}\bigg(\frac{\psi_i}{\varsigma_i}\bigg)\bigg],
\end{equation}

\begin{equation}
\label{tfi-fm}
\text{TFI} = \frac{[\sum_{i=1}^{u}{\mathcal{M}(\frac{\psi_i}{\varsigma_i})}]^2}{u\sum_{i=1}^{u}{{\mathcal{M}(\frac{\psi_i}{\varsigma_i})}^{2}}},
\end{equation}

\noindent where $\psi_i$ denote the throughput achieved for user $i$ at the 
Transport layer by the proposed scheduler, $\varsigma_i$ denote the Transport layer 
throughput received for user $i$ by the RR scheduler, $u$ is the total number of users and $\mathcal{M}$ is a positive real-valued function defined as:

\begin{equation}
\begin{split}
\label{wctfi-m}
\mathcal{M}(\vartheta)= \begin{cases} 
               \vartheta, & \text{if $0 \le \vartheta \le 1$} \\
	       1, & \text{otherwise}.
	\end{cases}
\end{split}
\end{equation}

Values of WCTFI and TFI varies between 0 and 1; 
0 for a completely unfair system and 1 for a completely fair system at the Transport Layer. 
Since the TCP-aware schedulers provide higher throughput than RR scheduler at all 
shadowing conditions (Fig. \ref{rr-tcp-aware-comp-4} and \ref{rr-tcp-aware-comp-3}), 
the values of both TFI and WCTFI of TCP-aware schedulers are 1. Therefore, TCP-aware 
schedulers are also Transport layer fair.

\subsubsection{Slot Utilization}
\begin{figure}[h!]
\centering
\includegraphics[width=0.53\textwidth]{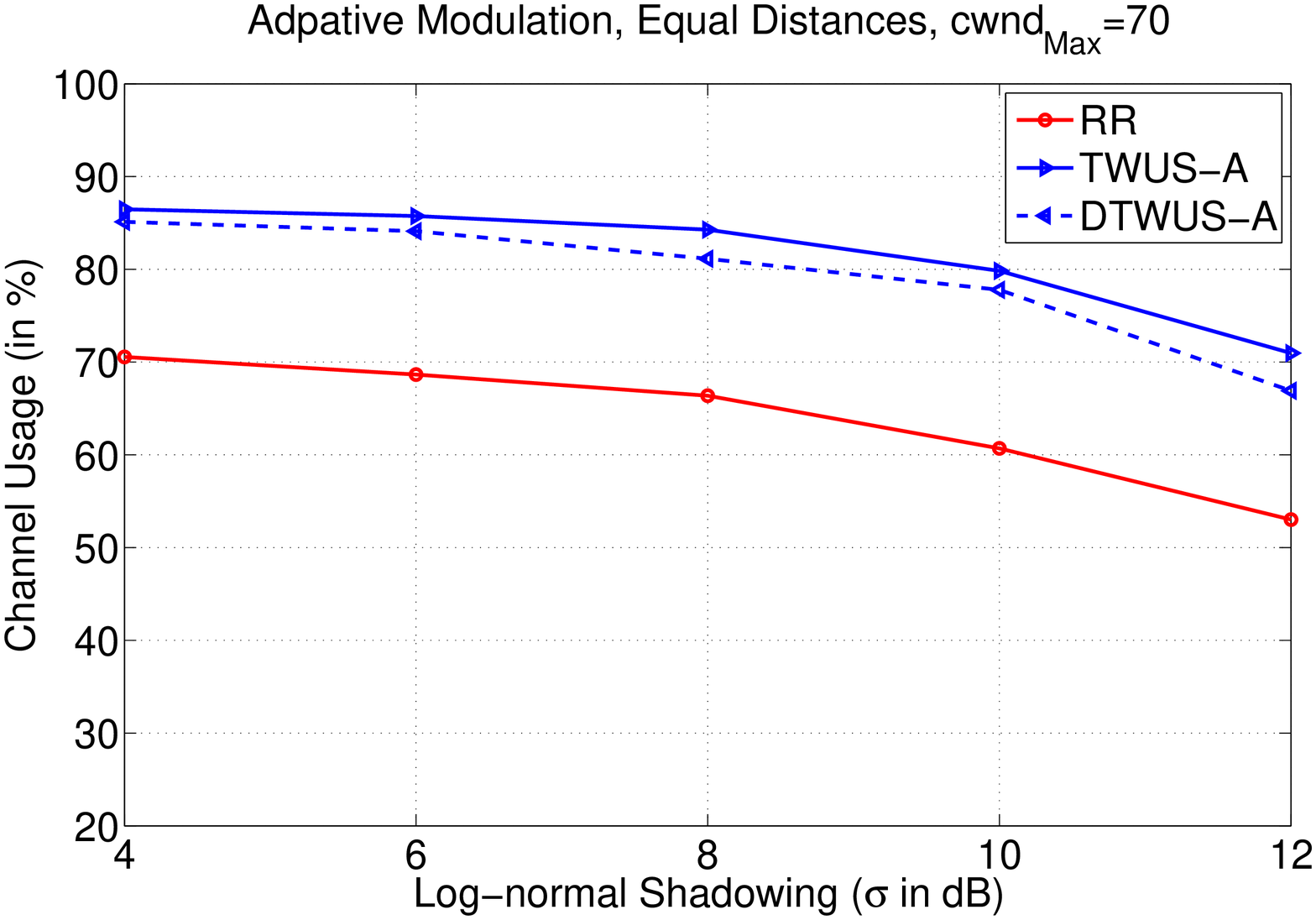}
\caption{Channel Utilization vs. Shadowing Parameter $\sigma$} 
\label{pf-tcp-aware-su-comp1}
\end{figure}

\begin{figure}[h!]
\centering
\includegraphics[width=0.53\textwidth]{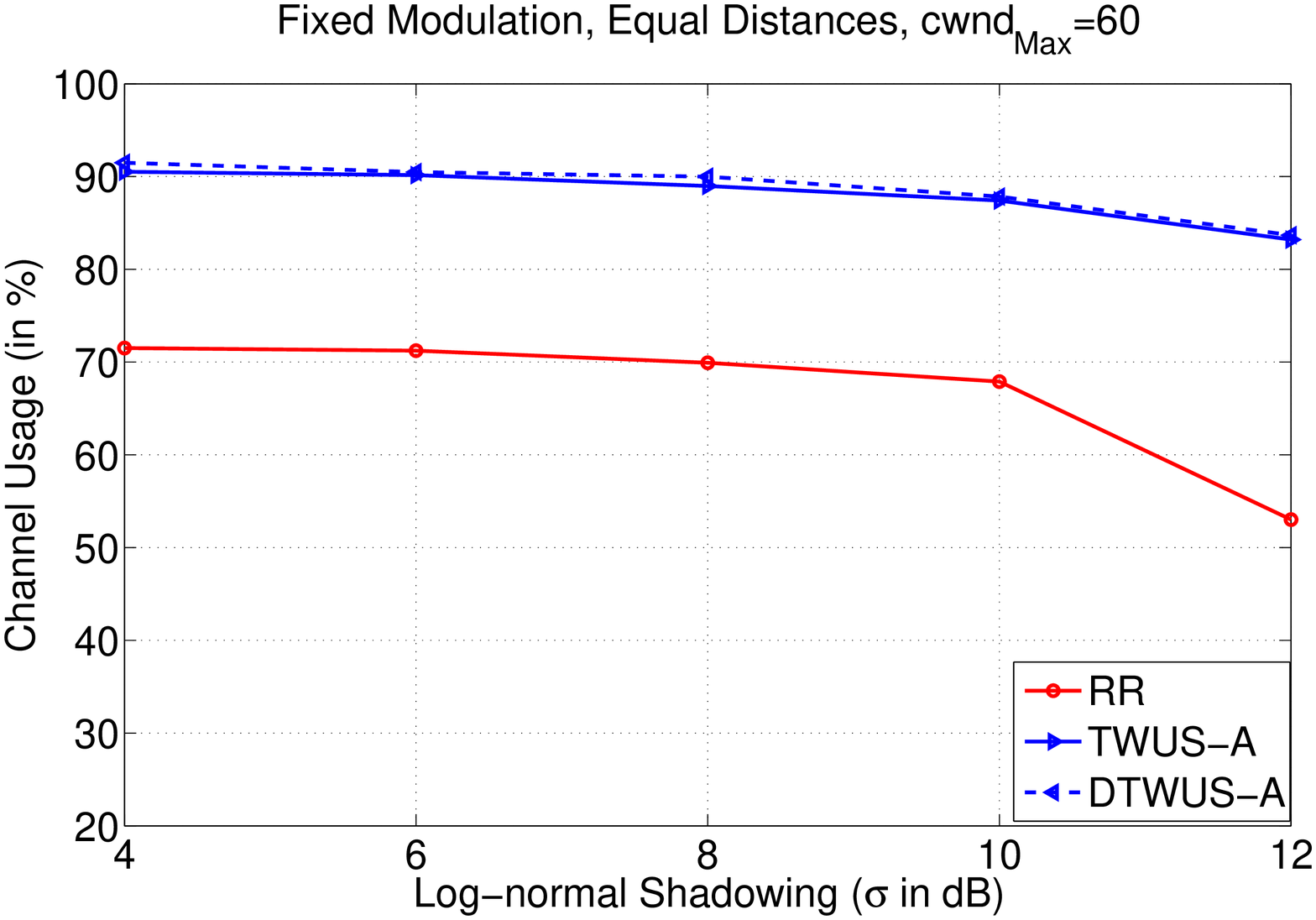}
\caption{Channel Utilization vs. Shadowing Parameter $\sigma$} 
\label{pf-tcp-aware-su-comp2}
\end{figure}

We also investigate the slot utilization of TWUS-A and DTWUS-A and compare it with RR 
scheduler. In Fig. \ref{pf-tcp-aware-su-comp1} and \ref{pf-tcp-aware-su-comp2}, we plot 
slot utilization of TCP-aware and RR scheduler with different value of $\sigma$. From these figures, we
observe that the slot utilization of TCP-aware schedulers is more than that of
RR scheduler. In addition, utilization of TWUS-A (TWUS) scheduler is more than that of DTWUS-A (DTWUS) scheduler. Note that even though TWUS-A has higher channel usage as compared to that of DTWUS-A (c.f., Fig. \ref{pf-tcp-aware-su-comp1}), throughput achieved by DTWUS-A scheduler is more than that of TWUS-A scheduler. This is due to the fact that the chance of TCP timeouts in DTWUS-A is lesser than TWUS-A and hence less retransmission of packets, resulting in higher throughput achieved (c.f., Fig. \ref{rr-tcp-aware-comp-4}).

We also observe that the slot utilization of TWUS-A and DTWUS-A scheduler varies between 
70\% to 85\%, whereas that of RR varies between 54\% to 70\%. 
Moreover, as the value of $\sigma$ increases, the slot utilization decreases. 
This is because, when the channel is under heavy shadowing, the probability of 
 not being scheduled in a frame is very high. This results in reduction in $cwnd$ size thereby 
resulting in low utilization. Similar results are 
also observed for unequal distance experiments and for fixed modulation experiments.


\subsection{Adaptive Modulation vs. Fixed Modulation} \label{comparison-expt}
From the simulation results presented in the previous section, 
we observe that the amount of data transmitted by the users
in adaptive modulation scheme is more as compared to the amount of data
transmitted in fixed modulation scheme. The average rate of transmission is
almost double (80-90\%) in adaptive modulation scheme as compared to fixed
modulation scheme for similar conditions. The higher rate of transmission in
adaptive modulation is achieved at the cost of extra complexity in the transmitter
and receiver structures at $SS$s. Adaptive modulation scheme also increases
average $cwnd$ size, resulting in higher TCP throughput. However, we observe that
schedulers with fixed modulation scheme have higher slot utilization than
schedulers with adaptive modulation scheme. The slot utilization in adaptive modulation schemes
can be increased by scheduling different classes of traffic along with TCP traffic. This needs
to be investigated further.

\section{TCP Throughput Analysis} \label{send_rate_ch5}
In \cite{padhye}, the authors have derived a closed form expression
for the steady state $send ~ rate$ of TCP flow as a function of loss rate, $RTT$
and TCP timeout. In this paper, we modify the 
expression of $send ~ rate$ in \cite{padhye} by incorporating the polling delay in the proposed scheduling.  
Let $p$ be the probability that $SNR_i \ge \min\{SNR_{th}\}$, for
any $SS_i$. Hence, the expected number of
polling epochs $L$ which a $SS$ needs to wait before becoming schedulable is,

\begin{equation}
\label{tcp-2}
E[L]=\sum_{L=1}^{\infty}{L p (1-p)^{L-1}} -1.
\end{equation}

The expected number of frames that a $SS$ waits is $E[L]\times k\times T_f$. 
This corresponds to  the average polling delay in the proposed scheduling algorithm. 
As discussed in \cite{padhye}, the average TCP $send ~ rate$ 
can be expressed as:
\begin{equation}
\label{tcp-6}
\begin{split}
B_{w} & \approx  \min \Bigg( \frac{cwnd_{Max}}{RTT_w}, \\ 
  &\frac{1}{RTT_{w}\sqrt{\frac{2bp_w}{3}} +
TO\min\left(1,3\sqrt{\frac{3bp_w}{8}}\right)p_w(1+32p_{w}^{2})}\Bigg),
\end{split}
\end{equation}

\noindent where $B_{w}$ is the TCP $send ~ rate$ or end-to-end throughput (in
packets per unit time), $b$ is the number of TCP packets acknowledged by one
ACK, $RTT_{w}$ is the average TCP round trip time, $p_w$ is the packet loss probability and $TO$ is the average
TCP timeout value. In TCP, since the congestion window size can grow
up-to $cwnd_{Max}$, maximum TCP $send ~ rate$ is bounded by
$\frac{cwnd_{Max}}{RTT_w}$. 

The end-to-end TCP throughput or $send ~ rate$ for our case
 can be modified by incorporating the polling delay into the round trip time. Accordingly,
$RTT_{w}$ in the above expression can be modified to include the polling delay in scheduling. The new round trip time
$RTT_{wr}$ is expressed as:

\begin{equation}
\begin{split}
\label{tcp-7}
RTT_{wr}& =  RTT_w+  E[L]\times k\times T_f.
\end{split}
\end{equation}

By replacing $RTT_{w}$ by $RTT_{wr}$ in (\ref{tcp-6}), TCP 
$send ~ rate$ for WiMAX network ($B_{wr}$) can be
expressed as:

\begin{equation}
\begin{split}
\label{tcp-8}
B_{wr} & \approx  \min \Bigg(\frac{cwnd_{Max}}{RTT_{wr}}, \\ 
&\frac{1}{RTT_{wr}\sqrt{\frac{2bp_w}{3}} +
TO\min\left(1,3\sqrt{\frac{3bp_w}{8}}\right)p_w(1+32p_{w}^{2})}\Bigg).
\end{split}
\end{equation}

\subsection{Validation of TCP Throughput}
We compare the average TCP $send ~ rate$ obtained in (\ref{tcp-8}) with our
simulation results. We determine the probability of loss ($p_w$) similar to that
of \cite{padhye}. We consider both triple-duplicate ACKs and TCP timeouts as
loss indications. Let $p_w$ be the ratio of the total number of loss indications
to the total number of packets transmitted. From simulations (with $\sigma$ of 
Log-normal shadowing as 8 dB and other parameters as shown in Table 
\ref{system_parameters-expt-a}), we observe that
the average probability ($p$) that $SNR_i \ge \min\{SNR_{th}\}$ is 0.87,
$\forall i \in I$. Using (\ref{tcp-7}), we determine $RTT_{wr}$. Then by
using the value of $RTT_{wr}$ obtained using (\ref{tcp-7}) and $p_w$
obtained above, we determine the average TCP $send ~ rate$ using
(\ref{tcp-8}). 

\begin{figure}[h!]
\centering
\includegraphics[width=0.53\textwidth]{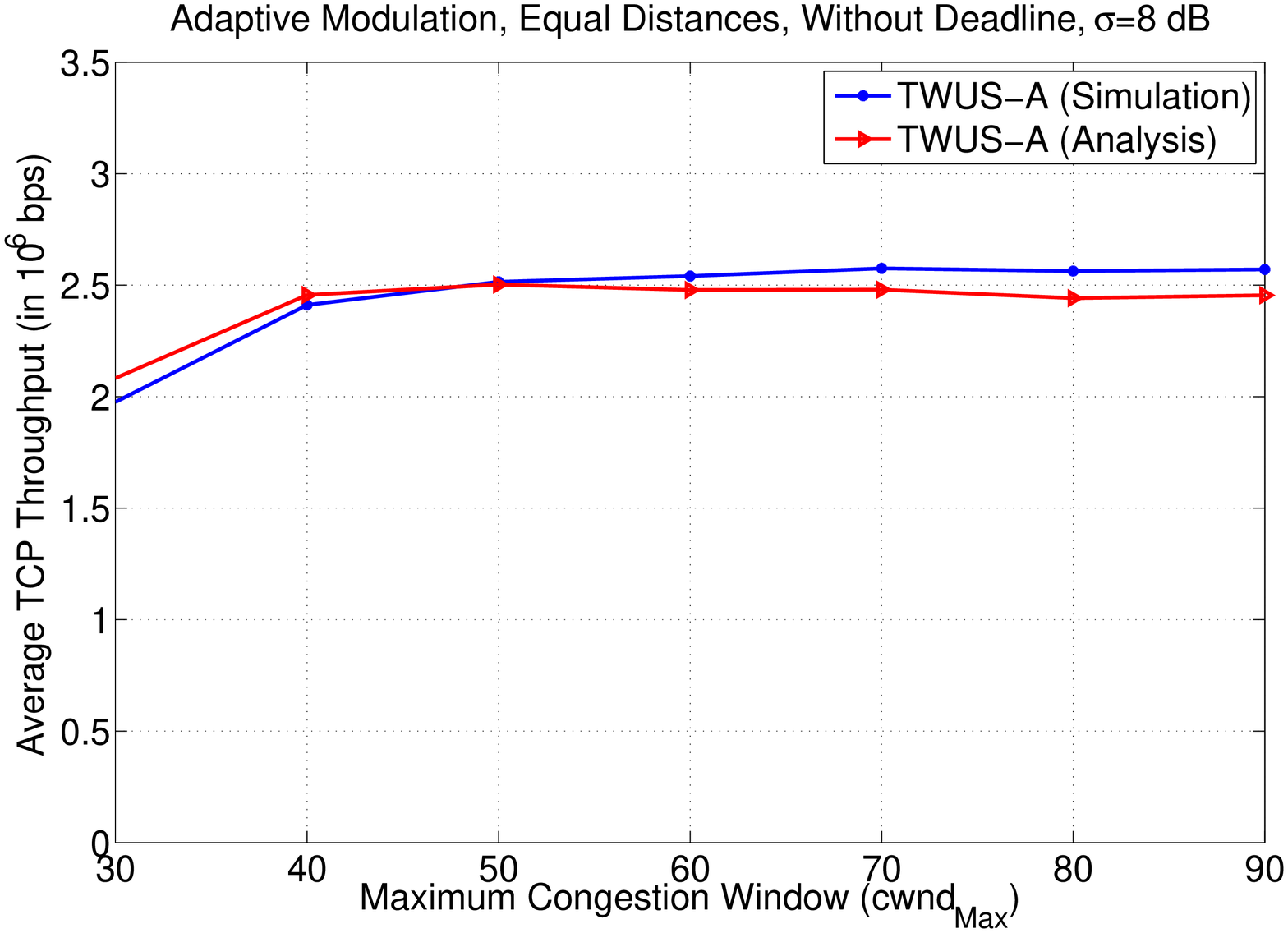}
\caption{Avg. TCP Throughput of TWUS-A at Different $cwnd_{Max}$}  \label{tcp-comp-tws5}
\end{figure}

\begin{figure}[h!]
\centering
\includegraphics[width=0.53\textwidth]{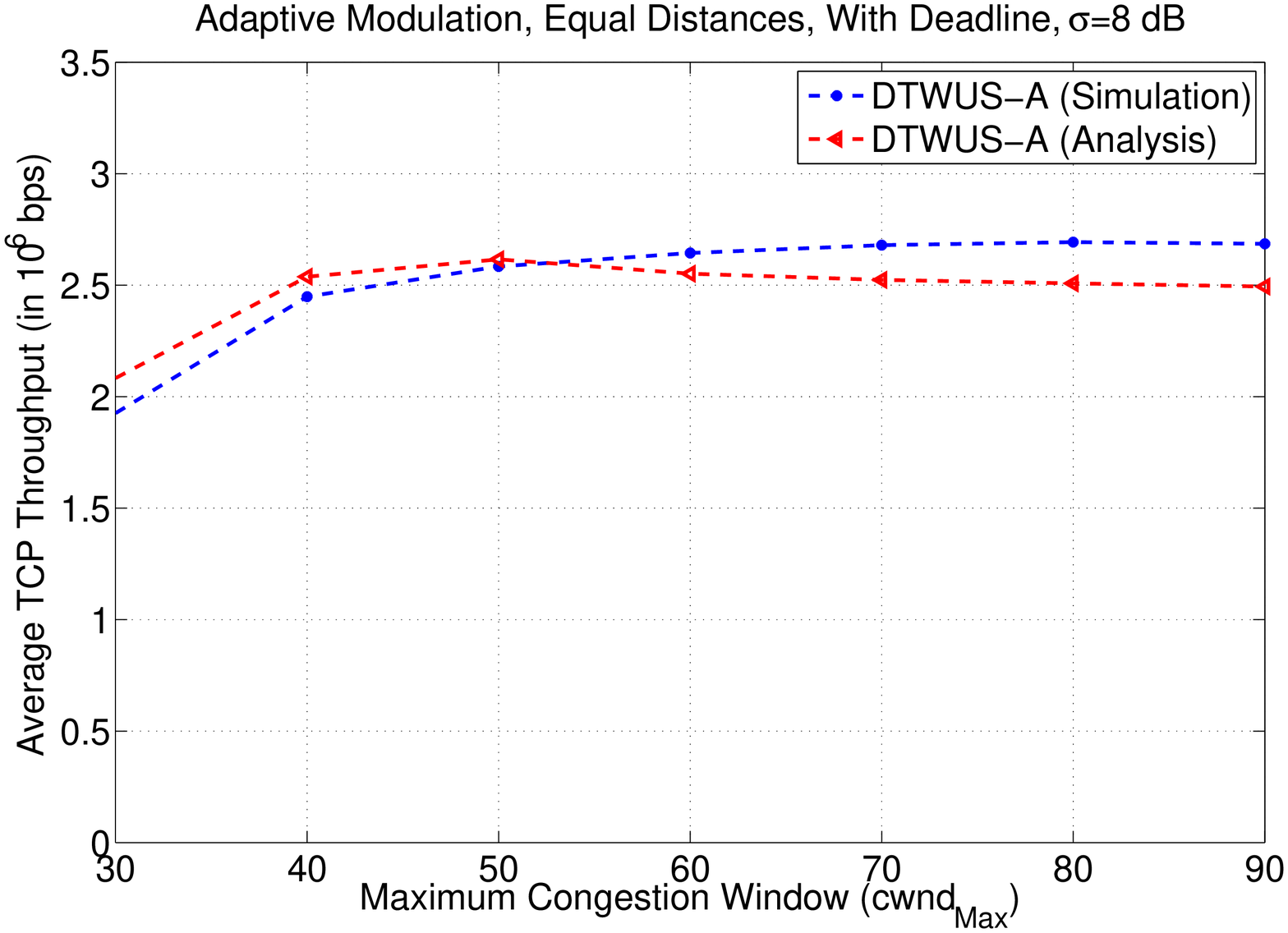}
\caption{Avg. TCP Throughput of DTWUS-A at Different $cwnd_{Max}$}
\label{tcp-comp-tws6} 
\end{figure}

\begin{figure}[h!]
\centering
\includegraphics[width=0.53\textwidth]{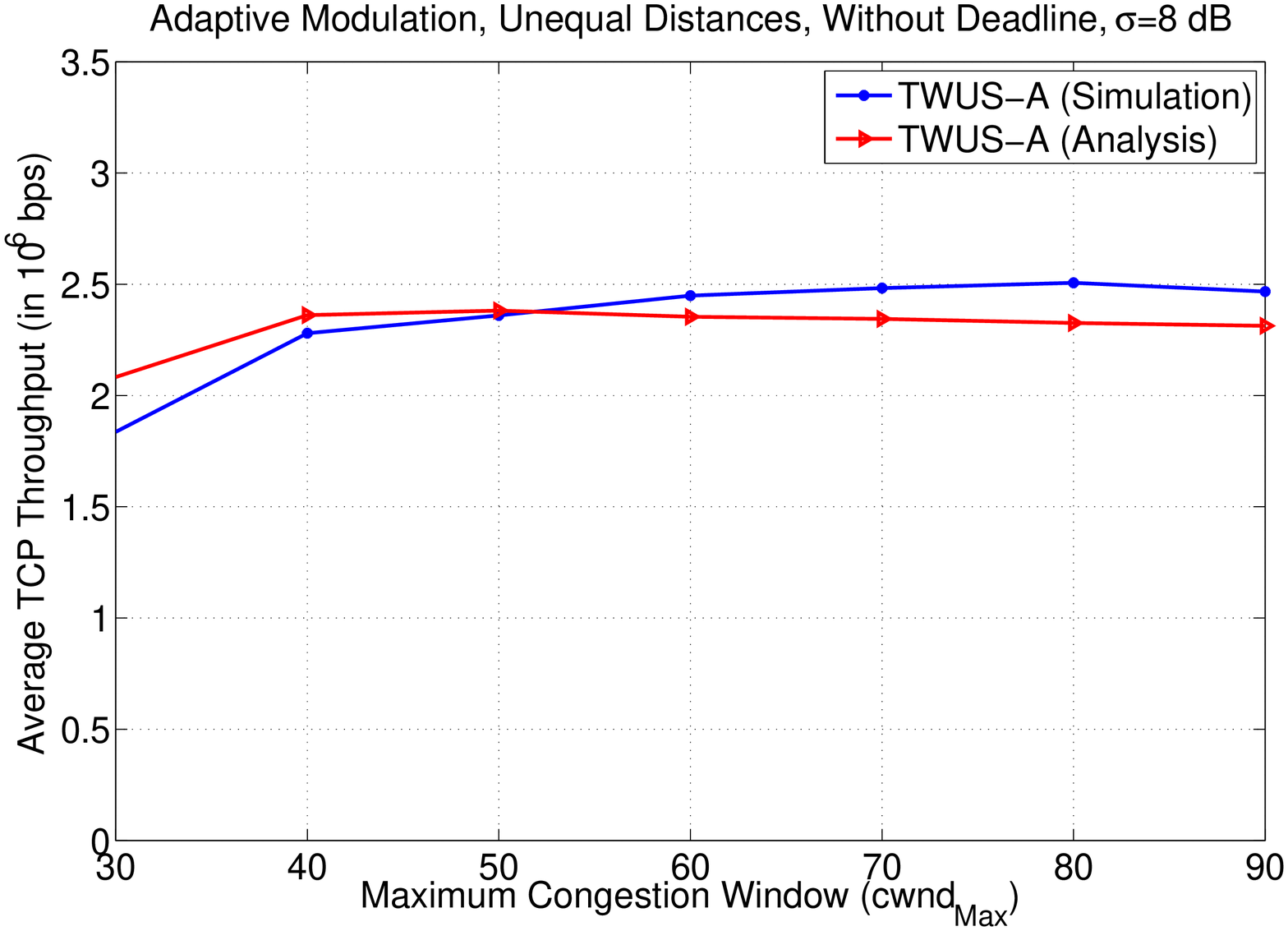}
\caption{Avg. TCP Throughput of TWUS-A at Different $cwnd_{Max}$} \label{tcp-comp-tws7} 
\end{figure}

\begin{figure}[h!]
\centering
\includegraphics[width=0.53\textwidth]{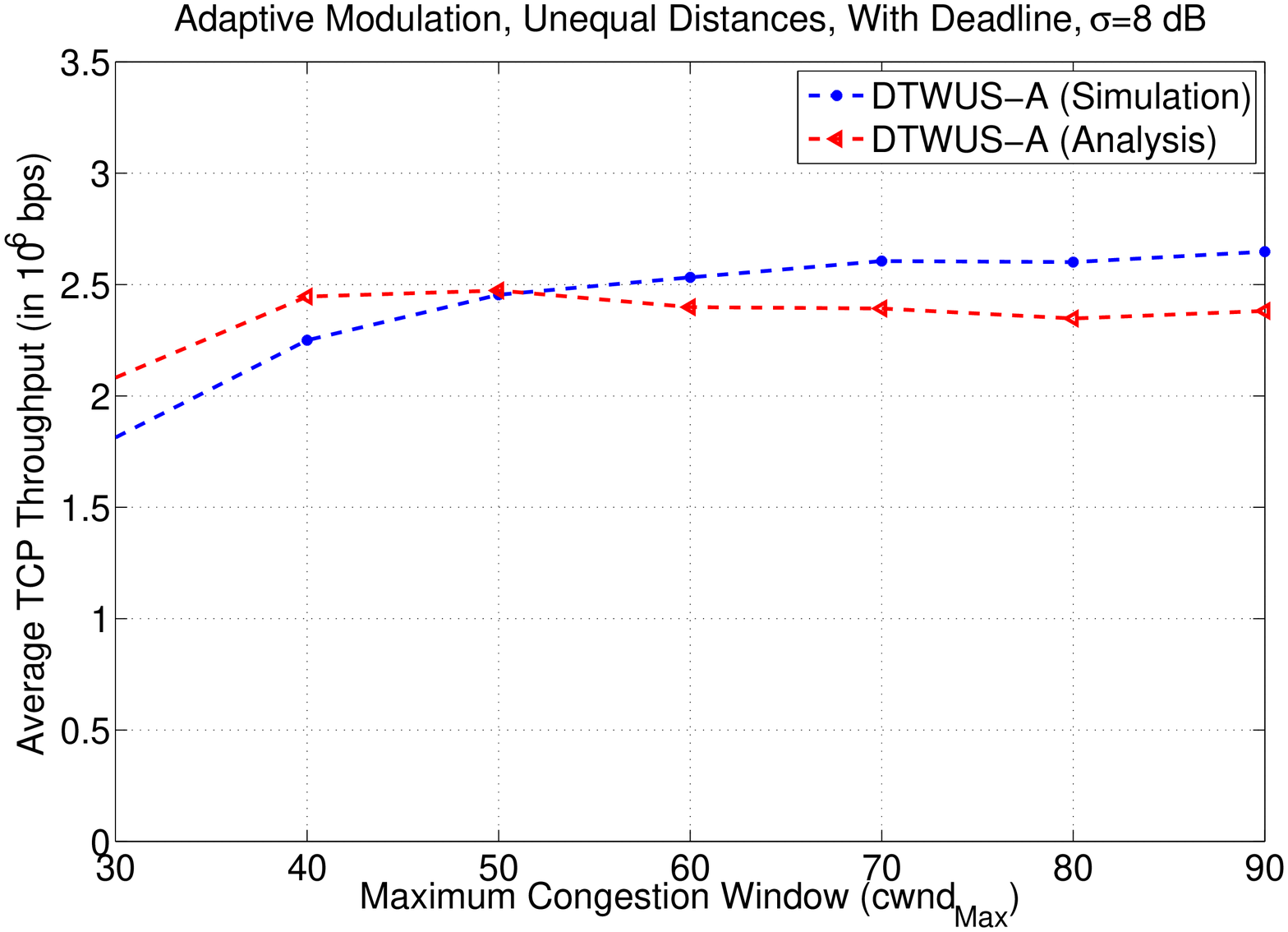}
\caption{Avg. TCP Throughput of DTWUS-A at Different $cwnd_{Max}$}\label{tcp-comp-tws8}  
\end{figure}

To verify our model, we determine average TCP $send ~ rate$ for all four sets of
experiments (TWUS-A with equal and unequal distances, DTWUS-A with equal and
unequal distances). We plot the analytical and experimental TCP throughput at
different $cwnd_{Max}$ in Fig. \ref{tcp-comp-tws5} - \ref{tcp-comp-tws8}.
From these figures, we observe that the theoretical $send ~ rate$s determined
using (\ref{tcp-8}) and the $send ~ rate$ obtained by our simulations 
match very closely. 

\section{Summary and Observations}\label{conclu_tcp_ch5}
In this paper, we have proposed scheduling algorithms for a multipoint-to-point network that adapts resource allocation based on TCP parameters-congestion window and timeout. The resource requirements are communicated during polling at {\it flow}
level. The scheduler also takes into account the wireless channel
characteristics and is thus cross layer in nature.
Further, we  have performed exhaustive simulations to investigate fairness and 
throughput behavior in WiMAX network setting. We have compared the performance of our scheduling schemes with that of RR scheduler under different shadowing. The proposed TCP-aware scheduling schemes perform better than RR scheduler in terms of slot utilization,
fairness and throughput.  

Though we have assumed that the downlink does not have any bandwidth constraint, in practice this assumption may not hold true. Hence, the effect of downlink congestion and the possible drop of $ACK$ packet on the TCP throughput needs to  be analyzed. In addition to this, as discussed before, the impact of scheduling of other class of traffic on scheduling of TCP traffic needs further investigation. Further, there is also a scope of extending this work for a high speed broadband mobile network such as IEEE 802.16m \cite{802.16m} based network. With mobility in place, scheduling of users to provide high data rates with hand-off 
margins is another area for investigation. 

\bibliographystyle{ieeetr}
\bibliography{thesis_hemant}
\end{document}